\documentclass[a4paper]{article}
\usepackage{multirow}
\usepackage{hyperref}

\usepackage{INTERSPEECH2020}

\title{Approaches to Improving Recognition of Underrepresented Named Entities in Hybrid ASR Systems}
\name{Tingzhi Mao$^1$, Yerbolat Khassanov$^{2,3}$, Van Tung Pham$^2$, Haihua Xu$^2$, Hao Huang$^1$, Eng Siong Chng$^2$}
\address{
$^1$School of Information Science and Engineering, Xinjiang University, Urumqi, China\\
$^2$School of Computer Science and Engineering, Nanyang Technological University, Singapore\\
$^3$ISSAI, Nazarbayev University, Kazakhstan
}
  
\email{tingzhimao@gmail.com, hwanghao@gmail.com, \{yerbolat002,vtpham,haihuaxu,aseschng\}@ntu.edu.sg}

\begin{document}

\maketitle
\begin{abstract}
In this paper, we present a series of complementary approaches to improve the recognition of underrepresented named entities (NE) in hybrid ASR systems without compromising overall word error rate performance.
The underrepresented words correspond to rare or out-of-vocabulary (OOV) words in the training data, and thereby can't be modeled reliably.
We begin with graphemic lexicon which allows to drop the necessity of phonetic models in hybrid ASR.
We study it under different settings and demonstrate its effectiveness in dealing with underrepresented NEs. 
Next, we study the impact of neural language model (LM) with letter-based features derived to handle infrequent words.
After that, we attempt to enrich representations of underrepresented NEs in pretrained neural LM by borrowing the embedding representations of rich-represented words.
This let us gain significant performance improvement on underrepresented NE recognition.
Finally, we boost the likelihood scores of utterances containing NEs in the word lattices rescored by neural LMs and gain further performance improvement.
The combination of the aforementioned approaches improves NE recognition by up to 42\% relatively.  
\end{abstract}
\section{Introduction}
The increasing popularity of voice-enabled services in mobile phones, smart homes and other internet-of-things devices triggered the resurgence of automatic speech recognition (ASR) technologies~\cite{luscher2019rwth,karita2019comparative,graves2014towards,chorowski2014end}.
As a result, the correct recognition of named entities (NE) such as street, restaurant and person names, to mention a few, has become of paramount importance for the proper functioning of these services~\cite{9054235}.
However, some of the NEs might be rare words appearing zero or a few times in the training set, and thereby leading to suboptimal ASR performance.
For example, Chinese, Malay and Indian NEs are frequently used in the context of Singapore English speech.
However, such NEs are extremely rare in general English speech corpora used to train Singapore English ASR.
Consequently, these rare NEs will be underrepresented, and thus, tend to be misrecognized by the ASR system causing the underlying applications such as natural language understanding (NLU) to fail.

In this work, we focus on hybrid deep neural network (DNN) hidden Markov model (HMM) based ASR systems which still achieve the state-of-the-art results as have been recently shown for various benchmarks~\cite{luscher2019rwth,karita2019comparative}.
Different from the newly proposed end-to-end (E2E) ASR architectures that operate at subword level~\cite{graves2014towards,chorowski2014end}, the hybrid DNN-HMM ASR systems mostly operate at word-level, and thus, are sensitive to both out-of-vocabulary (OOV) and rare words.
While the OOV problem has been extensively studied~\cite{egorova2018out,he2014subword,lin2007oov,thomas2019detection}, the rare words remain open research problem.

In hybrid ASR pipeline, there are several components which might cause the misrecognition of underrepresented words.
For example, grapheme-to-phoneme (G2P) model used to generate phoneme sequences of new words might be inaccurate; misleading the acoustic model.
Another reason might be the N-gram language model (LM) applied during the decoding stage which distributes most of the probability mass among frequent words.
Similarly, the neural LMs used during the rescoring stage also tend to be biased against both infrequent and out-of-shortlist\footnote{Words present in the vocabulary of ASR, but not in the vocabulary subset of neural LM.} words, since they might cover even smaller subset of frequent words present in the training data.
The cumulative effect of these components might be disastrous for underrepresented NEs. 

In this work, we aim to improve the recognition of underrepresented NEs in word-level hybrid DNN-HMM ASR without compromising overall WER performance. 
We study the cumulative effect of different existing approaches and propose several novel modifications.
Particularly, we first explore the context- and position-dependent graphemic lexicon~\cite{DBLP:conf/asru/LeZZFZS19} to remove the cost of learning G2P models.
Second, we study the effectiveness of neural LM with letter-based features~\cite{xu2018neural} used to rescore word lattices.
Next, we investigate the embedding matrix augmentation techniques~\cite{Khassanov2018,Khassanov2019} to enrich underrepresented NE representations in pre-trained neural LMs.
Lastly, we examine the effect of boosting the likelihood of utterances containing NEs in the word lattice.
We evaluated these approaches on Singapore English speech recognition task where we aimed to improve the recognition of rare Singapore related NEs.
The experiment results show that the combination of different approaches significantly improves the recognition accuracy of underrepresented NEs by up to 42\% relatively.
Importantly, all these approaches don't require additional labeled training data and don't degrade the overall WER performance. 

The rest of the paper is organized as follows.
In Section 2, we briefly review related works.
In Section 3, we explain our task in greater details.
In Section 4, we describe different approaches evaluated in this paper and present modifications contributed by us.
In Section 5, we first describe the experiment setup followed by results and analysis.
Lastly, Section 6 concludes this paper.

\section{Related works}
In hybrid DNN-HMM ASR, a common practice is to ignore the rare words by replacing them with the special $<$unk$>$ token.
In addition to reduced computation overhead, this will also reduce the potential conflicts between acoustically similar words leading to better overall WER.
Thus, this practice is preferable if the final goal is to reduce an overall WER where all words are of equal importance.
However, with the rise of voice-enabled services, the recognition of rare words is becoming crucial since they might be important proper nouns.

Unfortunately, the recognition of rare words is poorly studied as compared to the recognition of OOV words~\cite{egorova2018out,he2014subword,lin2007oov,thomas2019detection}.
To recognize OOV words, the early approaches employed hybrid word/subword recognition systems where $<$unk$>$ tokens are replaced by the sequence of subword units~\cite{szoke2008hybrid}.
Even though these approaches were designed to tackle OOV words, they can be also applied to recover rare in-vocabulary words~\cite{kombrink2010recovery}.
However, these approaches have several drawbacks such as recognition of non-existing words assembled by stacking different subword units.

Another group of similar tasks are information retrieval from speech~\cite{hatmi2013incorporating,jannet2017investigating,serdyuk2018towards,ghannay2018end}.
These tasks are either based on a pipeline of several systems or a single E2E system~\cite{serdyuk2018towards}.
The pipeline-based system usually consists of two parts: ASR and NLU.
The ASR is first used to transform speech into text which is then classified by NLU to structured data such as domain, intent and slots.
On the other hand, E2E systems aim to directly map speech into corresponding structured data. 
These tasks focus on improving the NE recognition and ignore the overall WER performance which is different from our goal.

Other existing approaches mostly focus on improving the recognition of rare words in LMs~\cite{xu2018neural,Khassanov2018,Khassanov2019,mikolov2012subword,kim2016character}.
For example,~\cite{xu2018neural,mikolov2012subword,kim2016character} proposed to employ hybrid word/subword tokens as input and output units of neural LMs.
On the other hand,~\cite{Khassanov2018,Khassanov2019} proposed to augment the representations of rare words in embedding matrices of pre-trained word-level neural LM.
These approaches are effective only if the rare words have been captured in the word lattices.

\section{Task description}\label{sec:td}
In this work, we aim to improve recognition of underrepresented NEs in a hybrid DNN-HMM ASR system.
In addition, we also want to preserve good overall WER performance which is different from previous works that mainly focus on NE extraction from speech.
To achieve our goal, in the following sections, we will present several complementary approaches and evaluate their impact using ablation study.
We define underrepresented NEs as infrequent proper nouns in training data.

The hybrid ASR is built to recognize Mandarin-English code-switching utterances.
It was trained using SEAME~\cite{lyu2010seame} speech corpus which is often employed for code-switching experiments~\cite{zzp2019,Khassanov2019C}.
The SEAME corpus consists of a train set and two test sets as shown in Table~\ref{tab:data}.
The test sets $\text{test}_{\text{man}}$ and $\text{test}_{\text{sge}}$ correspond to Mandarin English code-switching practices in Malaysia and Singapore, respectively. 
In addition, we used SG-streets\footnote{https://github.com/khassanoff/SG\_streets} dataset as an auxiliary evaluation set.
The SG-streets dataset consists of six recordings where Singaporean students read English passages about Singapore streets. 
To avoid ambiguity, the street names consisting of more than one word were joined using the underscore symbol, e.g. ‘boon lay’ is changed to ‘boon\_lay’.

To evaluate the underrepresented NE recognition capability of the hybrid ASR, we first extracted all NE words from SG-streets dataset whose frequency was less than ten in the train set.
The extracted NEs are mostly Singapore street names and we manually checked them to ensure their correctness.
In total, 393 underrepresented NEs were extracted and among them 195 don't appear in the train set at all, henceforth referred to as oov-NE.
The remaining 198 underrepresented NEs will be referred to as rare-NE.
Then, we computed WER on NEs recognized by the hybrid ASR.
In particular, we first align recognized hypotheses against references, and then collect segments that are aligned with the NEs.
Finally, we compute WER on these hypothesis-reference pairs which gives us NE-WER~\cite{garofolo2000trec}:
\begin{equation}\label{eq:ne-wer}
    \text{NE-WER}=\frac{\text{E}_{\text{NE}}}{\text{N}_{\text{NE}}}
\end{equation}
where $\text{E}_{\text{NE}}$ is the number of incorrectly transcribed NEs and $\text{N}_{\text{NE}}$ is the total number of NEs in reference.

To sum up, our final goal is to improve recognition of NE words from SG-streets dataset, while maintaining a good overall WER in all three evaluation sets. 
Thus, in our experiments, we will monitor both WER and NE-WER performances.
In the next section, we will present approaches employed to achieve our goal in greater details.

\begin{table}[t]
\caption{The overall dataset statistics}\label{tab:data}
\setlength{\tabcolsep}{2.0mm}
\centering
\begin{tabular}{l|c|c|c|c} 
\toprule
\multirow{2}{*}{}   & \multicolumn{3}{c|}{SEAME}                                            & \multirow{2}{*}{SG-streets} \\\cline{2-4}
                    & train     & $\text{test}_{\text{man}}$ & $\text{test}_{\text{sge}}$   &  \\ 
\midrule
\#Speakers           & 134       & 10                        & 10                        & 6 \\ 
Duration~(hours)    & 101.1     & 7.5                       & 3.9                       & 1.0 \\ 
\#Utterances         & 93,938    & 6,531                     & 5,321                     & 517 \\
\bottomrule
\end{tabular}
\end{table}

\section{Proposed approaches}
In this section, we describe different approaches employed to improve the recognition of underrepresented NEs.

\subsection{Graphemic lexicon}
Conventional hybrid ASR systems employ phonetic lexicons to map words to sequence of phonemes.
The phonetic lexicons are especially useful for languages where the correspondence between written form and pronunciation is poor as in the case of English.
Therefore, phonetic lexicons are carefully designed by linguists.
These lexicons can be further used to train G2P models to generate phoneme sequences of unseen words.
However, the performance of a G2P model might be suboptimal when it comes to difficult proper nouns such as Tchaikovsky.

To remove the dependency on G2P models, we evaluated the impact of graphemic lexicons on recognition of underrepresented NEs.
Specifically, we used recently proposed context- and position-dependent graphemic lexicon~\cite{DBLP:conf/asru/LeZZFZS19} which was shown to outperform phonetic lexicons on hybrid ASR systems.
We also studied its effectiveness when used with other approaches designed to deal with underrepresented words.

\subsection{Neural LM with letter-based features}
The neural LMs have become an indispensable component of ASR where it is used to rescore word lattices or N-best hypotheses list.
They help ASR systems to achieve the state-of-the-art results thanks to its exceptional generalization capability to unseen word sequences~\cite{luscher2019rwth,karita2019comparative}.
However, they can't learn reliable representations for infrequent words in the train set.
Consequently, during the rescoring stage, the neural LMs might push the hypotheses containing rare words to the bottom of the list.
To address this problem, the subword-level neural LMs have been extensively studied~\cite{xu2018neural,mikolov2012subword,kim2016character}.
In this work, we will evaluate the impact of neural LM with letter-based features~\cite{xu2018neural} on NE recognition.
The letter-based features complement neural LMs with subword level information, and thereby reinforce it against infrequent words.

\subsection{Embedding matrix augmentation}
Another approach to deal with the rare and OOV words in neural LMs is to apply embedding matrix augmentation techniques~\cite{Khassanov2018,Khassanov2019}.
The idea is to enrich the representations of infrequent words within pre-trained neural LM using the representations of similar and frequent words.
In this work, we further improved these techniques to achieve better results.
Specifically, we enrich the embedding representation of an underrepresented NE $e_u$ as follows:
\begin{equation}
\hat{e}_u = {\theta}{e_u} + \frac{\sum_{{e_c}\in{C_u}}{e_c}} {|C_u|}
\label{eq:enrich}
\end{equation}
where $\theta$ is a scaling factor, $C_u$ is a set of similar words, $e_c$ is an embedding representation of a similar word and $\hat{e}_u$ is an enriched representation of underrepresented NE.
The vocabulary enriched neural LM will be used to rescore word lattices.

\subsection{Word lattice boosting}
In keyword search (KWS) task, the main goal is retrieving the speech utterances containing a user-specified text query from a large database which is akin to our goal.
Therefore, we borrowed several approaches from KWS and further modified them for our task.
In particular, we first applied word lattice indexing technique~\cite{can2011lattice} to generate the inverted list of all words.
The inverted list contains the start-time, end-time and posterior probability scores of words.
Next, we boost the scores of all NEs while keeping scores of other words unchanged.
Lastly, the NE boosted list was used to regenerate a new word lattice from which the final 1-best output hypothesis is extracted.

\section{Experiment}\label{sec:exp}
In this section, we first describe the experiment setup followed by discussion and analysis of obtained results.

\subsection{Experiment setup} \label{sub:exp-setup}
All our experiments are conducted using hybrid DNN-HMM ASR built using the Kaldi toolkit~\cite{Povey2011TheKaldi}.
The ASR systems were evaluated using WER where word and character tokens were used for English and Mandarin, respectively.
The NE recognition performance was evaluated only on SG-streets set using NE-WER described in Eq.~\eqref{eq:ne-wer}.
The training and evaluation set characteristics are given in Table~\ref{tab:data}.

\textbf{Acoustic model:}
The acoustic model (AM) is built using the ‘nnet3+chain’ setup of Kaldi.
Specifically, our AM uses lattice-free maximum mutual information (LF-MMI)~\cite{Povey2016PurelySN} criterion over the factorized time-delay neural network (TDNN-F)~\cite{Povey2018SemiOrthogonalLM}.
The input features to AM are obtained by concatenating 43-dimensional MFFC features with pitch and 100-dimensional i-vectors.
The AM is composed of two networks: 6-layer convolutional neural network and 11-layer TDNN-F network with 1,536 units, bottleneck layers are set to 160 units.
In addition, we applied speed perturbation based data augmentation~\cite{ko2015audio}.

\textbf{Lexicon:}
The baseline lexicon was constructed by extracting all unique words from entire SEAME corpus.
Since the SEAME corpus consists of code-switching utterances, the lexicon is built by combining 13,685 English words and 2,660 Mandarin characters resulting in 16,345 unique tokens.
The phonetic lexicon for SEAME was carefully designed by linguists.
Note that phonetic lexicon size is slightly larger since some words have several pronunciation variants.
The OOV rate on $\text{test}_{\text{man}}$ and $\text{test}_{\text{sge}}$ is 0\%, while OOV rate on SG-streets is 11.47\%.

\textbf{Language model:}
All language models were trained on SEAME's train set transcripts.
During the decoding stage, we employed Kneser-Ney smoothed 4-gram LM (KN4) built using SRILM toolkit~\cite{stolcke2002srilm}.
During the rescoring stage, we employed neural LMs built using Kaldi-RNNLM~\cite{xu2018neural} interpolated with KN4\footnote{The interpolation weight of KN4 was set to 0.6.}.
It was trained as a 2-layer LSTM with 800 units in each layer.
The input and output embedding matrices were tied and set to 800 units.

\subsection{Experiment results and analysis} \label{sub:exp-analysis}
The performance of our baseline ASR system on SEAME's test sets is shown in Table~\ref{tab:baseline}.
We compared it against previous state-of-the-art systems to demonstrate that it achieves comparable results under similar training conditions.

\begin{table}[th]
\caption{The WER (\%) performance of our baseline ASR compared to previous works on SEAME test sets}\label{tab:baseline}
\renewcommand\arraystretch{1.1}
\centering
\begin{tabular}{l|c|c|c} 
\toprule
System                                      & LM rescoring  & $\text{test}_{\text{man}}$    & $\text{test}_{\text{sge}}$ \\ 
\midrule
E2E ASR~\cite{zzp2019}                       & Yes           & 25.0                          & 34.5 \\ 
hybrid ASR~\cite{guo2018study}               & No            & 22.8                          & 31.9 \\ 
hybrid ASR~\cite{maduo2019da}                & No            & 18.6                          & 25.4 \\\hline
hybrid ASR (ours)                            & No            & 18.6                          & 25.8 \\
\bottomrule
\end{tabular}
\end{table}

\begin{table*}[t]
\caption{WER (\%) and NE-WER (\%) performances for different ASR systems with phonetic (Ph) and graphemic (Gr) lexicons}
\label{tab:baseline wer}
\renewcommand\arraystretch{1.05}
\setlength{\tabcolsep}{1.5mm}
\centering
\begin{tabular}{c|l|c c|c c|c c|c c|c c|c c} 
\hline
\toprule
\multirow{4}{*}{No.} & \multirow{4}{*}{System}   & \multicolumn{6}{c|}{WER}  & \multicolumn{6}{c}{NE-WER} \\\cline{3-14}
                    &                           & \multicolumn{2}{c|}{\multirow{2}{*}{$\text{test}_{\text{man}}$}} &
\multicolumn{2}{c|}{\multirow{2}{*}{$\text{test}_{\text{sge}}$}} & \multicolumn{2}{c|}{\multirow{2}{*}{SG-streets}} & \multicolumn{6}{c}{SG-streets} \\\cline{9-14}
                    &                           &       &       &       &       &       &       & \multicolumn{2}{c|}{rare-NE}   & \multicolumn{2}{c|}{oov-NE} &  \multicolumn{2}{c}{ALL} \\\cline{3-14}
                    &                           & Ph    & Gr    & Ph    & Gr    & Ph    & Gr    & Ph    & Gr    & Ph    & Gr    & Ph    & Gr \\
\midrule
(1) & Baseline                                  & 18.6 & 18.7 & 25.8 & 26.0 & 39.1 & 39.0 & 14.7 & 15.2 & 100  & 100  & 57.0 & 57.3 \\
(2) & (1) + Expanded lexicon                    & 18.6 & 18.7 & 25.8 & 26.0 & 32.1 & 31.6 & 15.2 & 15.2 & 71.3 & 69.2 & 43.0 & 42.0 \\\hline\hline
(3) & (2) + Neural LM                           & \textbf{17.4} & 17.6 & \textbf{24.0} & 24.4 & 30.0 & 29.7 & 16.2 & 12.1 & 72.3 & 70.3 & 44.0 & 41.0 \\
(4) & (2) + Neural LM with letter-based features& 17.6 & 17.7 & 24.3 & 24.5 & 30.1 & 29.6 & 14.7 & 12.6 & 75.4 & 67.7 & 44.8 & 40.0 \\
\hline\hline
(5) & (3) + Embedding matrix augmentation       & 17.5 & 17.6 & 24.1 & 24.4 & 28.7 & \textbf{28.4} & 5.1 & 5.1  & 49.7 & 48.7 & 27.2 & 26.7 \\
(6) & (4) + Embedding matrix augmentation       & 17.6 & 17.7 & 24.3 & 24.5 & 28.9 & \textbf{28.4} & 5.6  & 5.1  & 52.3 & 48.7 & 28.8 & 26.7 \\
\hline\hline
(7) & (3) + Word lattice boosting               & --   & --   & --   & --   & 29.2 & 28.8 & 4.0  & 5.1  & 53.3 & 50.8 & 28.5 & 27.7 \\
(8) & (4) + Word lattice boosting               & --   & --   & --   & --   & 29.7 & 28.9 & 5.1  & 5.1  & 59.5 & 56.4 & 32.1 & 30.5 \\
(9) & (5) + Word lattice boosting              & --   & --   & --   & --   & 28.8 & 28.6 & \textbf{3.5} & 5.1  & 46.2 & \textbf{45.1} & \textbf{24.7} & 24.9 \\
(10) & (6) + Word lattice boosting               & --   & --   & --   & --   & 29.2 & 28.5 & \textbf{3.5} & 5.1  & 50.3 & 45.6 & 26.7 & 25.2 \\

\bottomrule
\end{tabular}
\end{table*}

\subsubsection{Graphemic lexicon}
The experiment results of baseline ASR using phonetic and graphemic lexicons are given in row (1) of Table~\ref{tab:baseline wer}.
The obtained results show that both lexicons achieve comparable WER and NE-WER on all evaluation sets.
Given that construction of graphemic lexicon is much easier, this observation is crucial for the development of hybrid ASR systems for new languages.

To further study the implications of graphemic lexicon, we expanded our baseline lexicon using OOV words from SG-streets set and re-build the baseline ASR.
In total, 506 unique OOV tokens were added, resulting in expanded lexicon covering 16,851 unique tokens.
To expand phonetic lexicon, we first trained G2P model using the baseline lexicon and then applied it to generate pronunciations for OOV words.
The graphemic lexicon was expanded by simply adding the written forms of OOV words.
The experiment results are given in row (2) of Table~\ref{tab:baseline wer}.

Expanding lexicons with OOV words improved WER on SG-streets set by around 7\% for both lexicon types with graphemic ASR achieving a slightly better result.
Additionally, the overall NE-WER performance of graphemic ASR improved over phonetic ASR by 1\%.
We hypothesize that the inferior performance of phonetic lexicon might be due to the error propagation from the G2P model.
This observation demonstrates the advantage of graphemic lexicon expansion and, in the future, further studies should be done in this direction.

\subsubsection{Neural LM with letter-based features}
Next, we evaluated the impact of word-level neural LM without and with letter-based features applied to rescore word lattices produced by the baseline ASR with expanded lexicon.
The vocabulary of neural LM was generated using expanded lexicon.
The letter-based features were generated by extracting 2-5 gram characters from the training set transcripts.
The experiment results are given in rows (3) and (4) of Table~\ref{tab:baseline wer}.

Compared to the system in row (2), rescoring with neural LMs further improves WER as anticipated for all evaluation sets.
However, the WER differences between neural LMs without and with letter-based features are marginal. 
The overall NE-WERs have slightly improved for both neural LMs on graphemic ASR, while for phonetic ASR they are degraded.
Tuning the hyper-parameters of neural LMs didn't help to achieve better results.
We conclude that using letter-based features to complement neural LM doesn't bring noticeable WER or NE-WER improvements\footnote{At least in our experiment setup.}.

\subsubsection{Embedding matrix augmentation}
In this experiment, we evaluated the embedding matrix augmentation technique applied to enrich representations of underrepresented NEs in pretrained neural LMs without and with letter-based features.
The representations were enriched using five frequent words, the interpolation weight $\theta$ in Eq.~\eqref{eq:enrich} was set to 0.01 and 0.09 for oov- and rare-NEs respectively.
The experiment results are given in rows (5) and (6) of Table~\ref{tab:baseline wer}.
Compared to systems in rows (3) and (4), the WER is unchanged for SEAME's test sets, while it is noticeably improved for SG-streets, especially for graphemic ASR.
Remarkably, the overall NE-WER has improved by 13.3\%-16.8\% for all neural LMs and lexicon types.
This observation demonstrates high efficacy of vocabulary enriching techniques for underrepresented NE recognition which is similar to what was reported in~\cite{Khassanov2018,Khassanov2019}.

\subsubsection{Word lattice boosting}
Lastly, we evaluated the word lattice boosting technique for SG-streets set.
We first applied it to word lattices rescored by standard neural LMs, i.e. rows (3) and (4).
The obtained results are given in rows (7) and (8) where small WER and large NE-WER improvements by 9.5\%-15.5\% are achieved.
We also applied it to word lattices rescored by vocabulary enriched neural LMs, i.e. rows (5) and (6), where additional NE-WER improvements by 1.5\%-2.5\% are achieved, see rows (9) and (10).

It is important to mention that the combination of different techniques such as neural LMs, vocabulary enriching and lattice boosting has substantially improved the recognition of rare-NEs, from 15.2\% down to 3.5\%.
The improvement in oov-NE recognition is also significant, from 69.2\% down to 45.1\%, but there is still room for improvement and further work is this direction should be done.
Another observation is that best NE-WER for rare-NEs is achieved on phonetic ASR, while best NE-WER for oov-NEs is achieved on graphemic ASR.

\section{Conclusions}
We investigated the efficacy of several complementary approaches to improve recognition of underrepresented NEs in hybrid ASR systems.
In particular, we considered phonetic and graphemic lexicons, neural LM with and without letter-based features, embedding matrix augmentation and word lattice boosting approaches.
We found that the cumulative effect of these approaches significantly improves the NE-WER from 43.0\% to 24.7\%, i.e. 42\% relative improvement. 
Importantly, they don't degrade the overall WER performance and don't require additional labelled data.
Interestingly, in our experiment setup, we observed that letter-based features are redundant as they don't help much the neural LMs to better recognize underrepresented NEs. 
Furthermore, we highlight that context- and position-dependent graphemic lexicons indeed achieve comparable results to phonetic lexicons, but much easier to construct and expand. 
We believe that observations reported in this paper will benefit many other applications where recognition of infrequent words is crucial and the construction of phonetic lexicon is difficult.

\bibliographystyle{IEEEtran}

\bibliography{mybib}

\end{document}